\documentstyle[12pt]{article}

\batchmode
  \newfont{\footscrfont}{rsfs10}
  \newfont{\footbbbfont}{msbm10}
\errorstopmode

\newif\ifscrf\scrftrue
\ifx\footscrfont\nullfont
  \scrffalse
\fi

\newif\ifamsf\amsftrue
\ifx\footbbbfont\nullfont
  \amsffalse
\fi

\def\ppnumber{\vbox{\baselineskip14pt\hbox{CLNS-95/1375}
\hbox{hep-th/9511171}}}
\def\ppdate{November 1995}
\def\pplogo{\vbox{\kern-\headheight\kern -15pt
\halign{##&##\hfil\cr&{
\ppnumber}\cr\rule{0pt}{2.5ex}&\ppdate\cr}
}}

\makeatletter
\date{}
\def\dedicatory#1{\def\@date{\normalsize\it#1}}
\def\subjclass#1{\def\@thefnmark{}\@footnotetext{1991
    {\it Mathematics Subject Classification.} #1}}
\def\keywords#1{\def\@thefnmark{}\@footnotetext{
    {\it Key words and phrases.} #1}}

\def\ps@firstpage{\ps@empty \def\@oddhead{\hss\pplogo}%
  \let\@evenhead\@oddhead 
}
\def\maketitle{\par
 \begingroup
 \def\thefootnote{\fnsymbol{footnote}}
 \def\@makefnmark{\hbox
 to 0pt{$^{\@thefnmark}$\hss}}
 \if@twocolumn
 \twocolumn[\@maketitle]
 \else \newpage
 \global\@topnum\z@ \@maketitle \fi\thispagestyle{firstpage}\@thanks
 \endgroup
 \setcounter{footnote}{0}
 \let\maketitle\relax
 \let\@maketitle\relax
 \gdef\@thanks{}\gdef\@author{}\gdef\@title{}\let\thanks\relax}

\def\abstract{\if@twocolumn
\section*{Abstract}
\else \small
\begin{center}
{\bf ABSTRACT}
\end{center}
\quotation
\fi}

\newif\iffn\fnfalse

\@ifundefined{reset@font}{\let\reset@font\empty}{} 
\long\def\@footnotetext#1{\insert\footins{\reset@font\footnotesize
    \interlinepenalty\interfootnotelinepenalty
    \splittopskip\footnotesep
    \splitmaxdepth \dp\strutbox \floatingpenalty \@MM
    \hsize\columnwidth \@parboxrestore
   \edef\@currentlabel{\csname p@footnote\endcsname\@thefnmark}\@makefntext
    {\rule{\z@}{\footnotesep}\ignorespaces
      \fntrue#1\fnfalse\strut}}}
\makeatother

\ifamsf
  \newfont{\bigbbbfont}{msbm10 scaled\magstep2}
  \newfont{\bbbfont}{msbm10 scaled\magstep1}  
  \newfont{\smallbbbfont}{msbm8}
  \newfont{\tinybbbfont}{msbm6}
  \newfont{\smallfootbbbfont}{msbm7}
  \newfont{\tinyfootbbbfont}{msbm5}
\fi

\ifscrf
  \newfont{\scrfont}{rsfs10 scaled\magstep1}  
  \newfont{\smallscrfont}{rsfs7}
  \newfont{\tinyscrfont}{rsfs7}
  \newfont{\smallfootscrfont}{rsfs7}
  \newfont{\tinyfootscrfont}{rsfs7}
\fi

\ifamsf
  \newcommand{\Bbb}[1]{\iffn
      \mathchoice{\mbox{\footbbbfont #1}}{\mbox{\footbbbfont #1}}
      {\mbox{\smallfootbbbfont #1}}{\mbox{\tinyfootbbbfont #1}}\else
      \mathchoice{\mbox{\bbbfont #1}}{\mbox{\bbbfont #1}}
      {\mbox{\smallbbbfont #1}}{\mbox{\tinybbbfont #1}}\fi}
\else
  \def\bigbbbfont{\bf}
  \def\Bbb{\bf}
\fi

\ifscrf
  \newcommand{\Scr}[1]{\iffn
    \mathchoice{\mbox{\footscrfont #1}}{\mbox{\footscrfont #1}}
    {\mbox{\smallfootscrfont #1}}{\mbox{\tinyfootscrfont #1}}\else
    \mathchoice{\mbox{\scrfont #1}}{\mbox{\scrfont #1}}
    {\mbox{\smallscrfont #1}}{\mbox{\tinyscrfont #1}}\fi}
\else
  \def\Scr{\cal}
\fi

\def\operatorname#1{\mathop{\rm #1}\nolimits}

\def\P{{\Bbb P}}

\def\R{{\Bbb R}}
\def\Z{{\Bbb Z}}

\def\Pic{\operatorname{Pic}}

\def\opeq#1{\advance\lineskip#1 \advance\baselineskip#1
	\advance\lineskiplimit#1}

\def\CY{Calabi-Yau}

\def\cM{{\Scr M}}

\def\cD{{\Scr D}}

\def\cMc{{\hfuzz=100cm\hbox to 0pt{$\;\overline{\phantom{X}}$}\cM}}
\def\barcD{{\hfuzz=100cm\hbox to 0pt{$\;\overline{\phantom{X}}$}\cD}}

\def\ff#1#2{{\textstyle\frac{#1}{#2}}}

\ifamsf

\else

\fi

\begin{document}
\setcounter{page}0
\title{\LARGE Enhanced Gauge Symmetries \\and Calabi--Yau Threefolds\\[10mm]}
\author{
Paul S. Aspinwall\\[0.7cm]
\normalsize F.R.~Newman Lab.~of Nuclear Studies,\\
\normalsize Cornell University,\\
\normalsize Ithaca, NY 14853\\[10mm]
}

{\hfuzz=10cm\maketitle}

\def\Large{\large}
\def\LARGE{\large\bf}

\vskip 1.5cm
\vskip 1cm

\begin{abstract}

We consider the general case of a type IIA string compactified on a
Calabi--Yau manifold which has a heterotic dual description. It is
shown that the nonabelian gauge symmetries which can appear
nonperturbatively in the type II string but which are understood
perturbatively in the
heterotic string are purely a result of string-string duality in six
dimensions. We illustrate this with some examples.

\end{abstract}

\vfil\break

\section{Introduction}		\label{s:intro}

Our understanding of the dynamics of $N=2$ Yang-Mills theories in four
dimensions has greatly improved recently due to the work of Seiberg
and Witten
\cite{SW:I,SW:II}. The moduli of such theories appear in two kinds of
supermultiplets, i.e., hypermultiplets and vector multiplets. Our
attention will focus on the vector multiplets in this letter. Consider
a theory with $n$ vector multiplets. This will correspond to a
Yang-Mills theory with a gauge group of rank $n$.

Roughly speaking, the vector moduli live in the Cartan subalgebra of
the gauge group and will break this group down to the elements that
commute with the value of the moduli. Thus, the effective gauge group,
$G$, will always be of rank $n$ and will generically be $U(1)^n$. At
particular points and subspaces within the moduli space the gauge group
will become nonabelian. This is the classical picture but quantum
effects modify this behaviour \cite{SW:I} (see also
\cite{KLTY:An,AF:An}). Instead of enhancement of the gauge group at
special points, the gauge group remains $U(1)^n$ but hypermultiplets
which are massive at generic points in the moduli space can become
massless. We want to consider the classical limit in which the gauge
group becomes enhanced but in the context of string theory. Having found
such points in the moduli space we can assume that Seiberg-Witten
theory will take over in the case of nonzero coupling.

Until recent developments in string duality it was generally believed
that a type II superstring compactified on a \CY\ manifold would not
exhibit a nonabelian gauge symmetry. When it was realized that the
type IIA string compactified on a K3 surface could naturally be
identified as the dual of the heterotic string on a four-torus
\cite{HT:unity,W:dyn,HS:sol} then it followed that, since the heterotic
string can have nonabelian gauge symmetries, the same must be true for
the type IIA string on a K3 surface. The way in which this was
possible was discussed in \cite{W:dyn} and was further developed in
\cite{me:enhg,W:dyn2,BSV:D-man}.

A similar duality in four dimensions has been conjectured
\cite{KV:N=2,FHSV:N=2} in which the type II string compactified on a
\CY\ threefold is considered equivalent by some kind of duality to a
heterotic string compactified on K3$\times T^2$ (or some variant
thereof). Since this heterotic string can again lead to nonabelian
gauge groups the same must be true of type II strings on a \CY\
threefold. Based on these conjectured dualities, points in the moduli
space for specific examples of compactification of type II strings on
\CY\ manifolds where the gauge group should become enhanced were identified.
In these cases the reduction to Seiberg-Witten theory for nonzero
coupling has been explicitly shown in \cite{KKL:limit}.

Our goal here will be to make general statements about the appearance
of enhanced gauge groups in type IIA compactifications.
The basic idea is actually very simple thanks to the results of
\cite{AL:ubiq}. We will restrict our attention (until some comments at
the end) to gauge groups which are visible perturbatively in the
heterotic string picture. The dual picture to this is a type IIA
string theory compactified on $X$ which must be a K3 fibration and the
gauge group arises from properties of the generic fibre. It was
shown that the dilaton modulus in the heterotic string corresponds to
the size of the base $\P^1$ of the fibration and that the
weak-coupling limit corresponds to the size of this base space being
taken to infinity. Clearly if we look at a generic fibre and the base
space becomes infinitely large then we have effectively decompactified
the 4-dimensional picture to 6 dimensions and we have reduced the
question the that studied in \cite{W:dyn,me:enhg}.

In a recent paper \cite{BSV:D-man} some questions regarding enhanced
gauge groups were studied using $D$-branes, particularly in the type
IIB context. Our results here concern type IIA strings but clearly
have some overlap with the conjectures and examples of
\cite{BSV:D-man}. The $D$-brane approach is probably the most powerful
tool for answering questions regarding type II string
compactifications but the purpose of this letter is to demonstrate how
simply many of the properties can be derived from what we already know
about 6-dimensional duality.

In section \ref{s:K3f} we will review the machinery we need to apply
our picture to examples and in section \ref{s:egs} we give
some examples. Finally we present some concluding remarks in section
\ref{s:conc}.


\section{K3 Fibrations}		\label{s:K3f}

We will be concerned with the case that $X$ can be written as a
fibration over $\P^1$ with generic fibre a K3 surface. The importance
of such a class in the context of string duality was first realized in
\cite{KLM:K3f,VW:pairs}.
Let us consider the general case of a dual pair consisting of a type
IIA string compactified on a \CY\ manifold, $X$, and a heterotic string such
that the weakly-coupled heterotic string is identified with $X$ at
some kind of large radius limit. It was shown in \cite{AL:ubiq} that
$X$ must be a K3-fibration in this case.

For a generic dual pair, the gauge group is $U(1)^{r+1}$. This comes
from $r$ vector multiplets and the graviphoton. In the type IIA
picture, the vector multiplets come from $H^{1,1}(X)$ and
$r=h^{1,1}(X)$. The group $H^{1,1}(X)\cap H^2(X,\Z)$ is known as the
Picard group of $X$. The rank of the Picard group is the Picard
number. In our case we will assume that $h^{2,0}(X)=0$
and so the Picard group of $X$ is simply $H^2(X,\Z)$. That is, the
generators of the Picard group of $X$ can be considered to form a basis of
the space of vector multiplets.

The Picard group of $X$ when $X$ is a K3 fibration essentially has
three different contributions. We work in terms of the dual group
$H_4(X)$. The generic K3 fibre itself gives one contribution. Another
source is from elements of the Picard group of the K3 fibre. In this
case a curve in the fibre is transported over the whole base $\P^1$ to
build up a 4-cycle. Note that such curves may have monodromy under
such a transport. One can see that monodromy-invariant elements of the
Picard group of the generic K3 fibre contribute to the Picard group of
$X$. Note that the Picard group of a K3 surface is a more subtle
object than the Picard group of a \CY\ threefold. The Picard group of
a K3 surface can depend upon the complex structure of the K3 surface
since $h^{2,0}\neq0$.
Lastly, there will be degenerate fibres over a finite number of points
in the base $\P^1$. In some cases such fibres can also contribute to
the Picard group of $X$.

In \cite{AL:ubiq} these three sources for elements of the Picard group
of $X$ were given different interpretations in terms of the dual
heterotic string. The generic fibre element was identified with the
dilaton, which lies in a vector multiplet. The elements coming from
the Picard group of the fibre were matched with the rest of the gauge
group that was visible perturbatively in the heterotic string. Lastly,
the contributions from the degenerate fibres were expected to have
some nonperturbative origin in the heterotic string.

We are interested in the gauge group that is perturbatively visible in
the heterotic string --- that is the part associated to the Picard
group of the K3 fibres. We would like to know if we can obtain
nonabelian groups by varying the moduli in these vector multiplets in
the weakly-coupled limit. The weakly coupled limit corresponds to the
base $\P^1$ becoming very large. Clearly in this limit, any question
about the generic fibres can be treated purely in terms of the type
IIA string compactified on a K3 surface along the lines of
\cite{W:dyn,me:enhg}.

This means that the analysis is actually very simple --- the question
of enhanced groups for $N=2$ theories in four dimensions is actually
completely reducible to the case of six-dimensions, at least as far as
the perturbative heterotic string is concerned.

To simplify our
discussion we will make an assumption concerning the way in which $X$
is written as a K3 fibration. We will assert that the Picard lattice
of a generic fibre is invariant, i.e., does not undergo any monodromy
transformation, as we move about the base space. One can certainly
find examples which will not obey our assumption but simple examples,
such as the ones we discuss later, do not have monodromy. It should
not be difficult
to extend the analysis here to nontrivial monodromy.

We are thus concerned with the enhanced gauge groups that can appear
on a K3 surface as we vary its K\"ahler form. The fibration of $X$
will restrict the K3 fibre to be of a particular type. Indeed, the generic
fibre will be an algebraic K3 surface which may be considered to be
embedded in some higher-dimensional projective space. An abstract K3
surface has variations in its complex structure and K\"ahler structure
which naturally fill out a space of 80 real dimensions in string
theory \cite{Sei:K3}. In the case of a specific algebraic K3 surface
however only some of these deformations are allowed. In particular,
the K\"ahler form is only allowed to vary within the space spanned by
the Picard group.

For the general case, the stringy moduli space of K3 surfaces is
given locally by the Grassmanian of space-like 4-planes in $\R^{4,20}$. The
space $\R^{4,20}$ may be viewed as the space of total real cohomology
$H^*({\rm K3},\R)$ which contains the lattice $H^*({\rm K3},\Z)$
which has intersection form $E_8\oplus E_8\oplus H\oplus H\oplus
H\oplus H$ \cite{AM:K3p}. Here $E_8$ denotes {\em minus\/} the
Cartan matrix of the Lie group $E_8$ and $H$ is the hyperbolic
plane. The global form of the moduli space is obtained by dividing
this Grassmanian by the group of isometries of this lattice.

In the case of an algebraic K3 surface, this moduli space is
restricted as follows.\footnote{This analysis rests heavily on work
done in collaboration with D.~Morrison \cite{AM:K3m}. Aspects have
also been discussed in \cite{Mart:CCC,Dol:K3m}.} Divide the
lattice as
\begin{equation}
  H^*({\rm K3},\Z) \supseteq \Lambda_K \oplus \Lambda_c.
		\label{eq:decomp}
\end{equation}
That is, take $\Lambda_K$ to be a sublattice of the integral
cohomology of the K3 surface and $\Lambda_c$ to be its orthogonal
complement. We demand that the sum of the ranks of these lattices is
equal to the rank of $H^*({\rm K3},\Z)$. That is, the lattice
$\Lambda_K \oplus \Lambda_c$ is equal to $H^*({\rm K3},\Z)$ or is a
sublattice of finite index.

Let $V_K=\Lambda_K\otimes_\Z\R$ be the real vector spanned by the
generators of $\Lambda_K$ with $V_c$ similarly defined. Now we may
consider the Grassmanian of space-like 2-planes in $V_K$ to be our
restricted moduli space of complexified K\"ahler forms and the
Grassmanian of space-like 2-planes in $V_c$ to be the restricted
moduli space of complex structures. Clearly the product of these two
moduli spaces is a subspace of the total stringy moduli space of K3
surfaces locally. Globally some of the isometries of $H^*({\rm
K3},\Z)$ will descend to isometries of $\Lambda_K$ and $\Lambda_c$ to
give identifications within these Grassmanians.

By restricting to a special class of K3 surfaces we have thus managed
to locally factorize the moduli space into deformations of complex
structure and deformations of K\"ahler form. If we want to make contact with
classical geometry then we must insist that the moduli space of
K\"ahler forms contains the large radius limit. In this case
\begin{equation}
  \Lambda_K = \Pic{} \oplus H,	\label{eq:LK}
\end{equation}
where Pic is the Picard lattice of the algebraic K3 surface. The $H$
factor is then identified with $H^0\oplus H^4$.

Let us denote the Picard number of the fibre by $r$. Since the
signature of the Picard lattice is $(1,r-1)$, we see immediately from
(\ref{eq:LK}) that the moduli space of K\"ahler forms within the fibre
is given by
\begin{equation}
  \frac{O(2,r)}{O(2)\times O(r)},
\end{equation}
divided by the group of isometries of $\Lambda_K$. Thus we give a
geometrical interpretation to the result of \cite{FvP:Ka}.

Note that the mirror map exchanges the r\^oles of $\Lambda_K$ and
$\Lambda_c$. Thus the mirror of one algebraic K3 surface is generally a
different algebraic K3 surface and the Picard numbers of these two
surfaces will add up to 20.

Now recall how, using string-string duality in six dimensions, we find
the points in the moduli space of type IIA strings on a K3 surface
where we have enhanced gauge groups \cite{W:dyn,me:enhg}. Let $\Pi$ be
the space-like 4-plane in $H^*({\rm K3},\R)$. The set of vectors
\begin{equation}
  \{\alpha\in H^*({\rm K3},\Z); \alpha^2=-2 \;{\rm and}\;
\alpha\in\Pi^\perp\},
\end{equation}
give the roots of the semi-simple part of the gauge group $G$.

For the case of interest to us, we are only concerned with the slice
of the moduli space given by deformations of the K\"ahler form on the
K3 surface. Given the decomposition (\ref{eq:decomp}) we see that our
gauge group now has roots
\begin{equation}
  \{\alpha\in \Lambda_K; \alpha^2=-2 \;{\rm and}\;
\alpha\in\mho^\perp\},	\label{eq:xG}
\end{equation}
where $\mho$ is the space-like 2-plane\footnote{The reason for this
notation comes from \cite{Cand:mir} as explained in \cite{AM:K3p}.} in
$V_K$.

This essentially gives the full description of how the generic fibres
can enhance the gauge group in the limit that the base space becomes
infinitely large, i.e., when the dual heterotic string becomes
weakly-coupled. First find the Picard group of the generic fibre, then
build $\Lambda_K$ from (\ref{eq:LK}). The gauge group is then given by
(\ref{eq:xG}).


\section{Examples}  \label{s:egs}

Let us clarify the discussion of the last section by giving two
examples. The first example we take from \cite{KV:N=2}. Let $X_0$ be the
\CY\ hypersurface of degree 24 in the weighted projective space
$\P^4_{\{1,1,2,8,12\}}$. The first
two homogeneous coordinates may be used as the homogeneous coordinates
of the base $\P^1$ to form a K3-fibration. The generic fibre may then
be written as a degree 12 hypersurface in $\P^3_{\{1,1,4,6\}}$ which
is indeed a K3 surface which we denote $F_0$.

The space $\P^3_{\{1,1,4,6\}}$ contains a curve of $\Z_2$-quotient
singularities which $F_0$ intersects once at a point. We blow-up the quotient
singularities in $X_0$ to obtain $X$ and this induces a blow-up of
$F_0$ which we call $F$. The latter is locally a blow-up of a $\Z_2$-quotient
singularity and so gives us a rational curve of self-intersection $-2$ within
$F$. Let us call the homology class of this curve $C$. Clearly $C$
lies in the Picard group of $F$. The other contribution to the Picard group
comes from hyperplanes of $\P^3_{\{1,1,4,6\}}$ slicing $F_0$. Let us
denote the resulting curve $A$. Representatives of $A$ pass through
the quotient singularity of $F_0$. When $F_0$ is blown-up such
self-intersections are removed and thus $A$ has self-intersection 0 in
$F$. Since $A$ passed through this singularity in $F_0$, the
intersection number between $A$ and $C$ is equal to 1.

We claim then that $F$ has Picard number 2 with intersection lattice
\begin{equation}
\left(\begin{array}{cc} 0&\phantom{-}1\\1&-2 \end{array}\right).
\end{equation}
Clearly by replacing $C$ with the cycle $B=A+C$ we obtain an
intersection form between $A$ and $B$ equal to $H$, the hyperbolic
plane.

Thus we obtain
\begin{equation}
  \Lambda_K \cong H\oplus H
\end{equation}
and the part of the moduli space coming from the fibre is given by
\begin{equation}
  \frac{O(2,2)}{O(2)\times O(2)}
\end{equation}
divided by $O(2,2;\Z)$ in the usual language. This is, of course, the
Narain moduli space for a string on the 2-torus. If the conjecture in
\cite{KV:N=2} is true then this is no accident as explained in
\cite{VW:pairs}. This part of the moduli space would arise from
deformations of the $T^2$ of the K3$\times T^2$ on which the dual
heterotic string is compactified.

Now we can look for places in the moduli space where the gauge group
is enhanced. The easiest case is when the K3 fibre can be taken to be
at large radius. Thus corresponds to a direction in $\mho$ becoming
almost light-like being close to a null vector in the $H$ factor in
(\ref{eq:LK}) \cite{AM:K3p}. Consistent with such a limit be may take
$\mho$ to be perpendicular to $C$. This means that the K\"ahler form
or $B$-field when integrated over the curve $C$ is zero. That is, we
have blown down $F$ back to $F_0$. The roots corresponding to $\pm C$
give the root lattice of $SU(2)$. Thus we expect an $SU(2)$ gauge
symmetry on the space $X_0$ (for suitable $B$-field).

$X_0$ contains a curve of $\Z_2$-quotient singularities. Note that
this is easy to generalize --- when we can enhance the gauge symmetry
by blowing down a curve in the K3 generic fibre then every fibre will
have a singular point meaning that $X$ will contain a curve of
singularities. The fact that curves of singularities can be associated
with enhanced gauge symmetries in a type IIB context has been
discussed in \cite{BSV:D-man}.

We can also embed the root lattice of $SU(2)\times SU(2)$ or $SU(3)$
into $\Lambda_K$ for the case at hand. In both cases the plane $\mho$
is fixed and so these are isolated points in the moduli space. Note
also that in both cases we cannot be near the large radius limit of
the K3 fibre. These further enhancements of the gauge group thus
correspond to effects of quantum geometry within the K3 fibre when the
volume of the fibre will be of the order of $(\alpha^\prime)^2$.

We can make closer contact with the conjecture of \cite{KV:N=2} by
going to the mirror picture of a type IIB string compactified on $Y$,
the mirror of $X$. In this case, $Y$ is an orbifold of the
hypersurface
\begin{equation}
  x_1^{24}+x_2^{24}+x_3^{12}+x_4^3+x_5^2+a_0x_1x_2x_3x_4x_5
  +a_1x_1^6x_2^6x_3^6+a_2x_1^{12}x_2^{12}=0
\end{equation}
in $\P^4_{\{1,1,2,8,12\}}$. Application of the monomial-divisor mirror
map in the \CY\ phase as described in \cite{AGM:sd} immediately tells
us that, to leading order, the size of the base $\P^1$ is given by
$\log(a_2^2)$; the size of the blown-up curve, $C$, within the K3
fibre is $\log(a_1^2/a_2)$; and the size of the fibre itself is given
by $\log(a_0^6/a_1)$.
Thus, the weak-coupling limit of the heterotic
string is given by $a_2\to\infty$ from \cite{AL:ubiq} in agreement
with the conjecture of \cite{KV:N=2}.

Keeping the K3 fibre big, by keeping $a_0^6/a_1$ big, we can blow down
$C$ by decreasing $a_1^2/a_2$. In the limit that $a_1^2/a_2$ becomes
zero we reach the conformal field theory orbifold. This is not what we
want however. In order to get the enhanced gauge symmetry we need the
component of $B$ along the blow-up to be zero as explained in
\cite{me:enhg}, whereas the conformal field theory orbifold gives a
value of $\ff12$. Fortunately for this blow-up, the $B$-field value for the
mirror map has been explicitly worked out in section 5.5 of
\cite{AGM:sd}. To obtain a zero-sized exceptional divisor and zero
$B$-field we require $a_1^2/a_2=4$. Thus we expect an $SU(2)$ enhanced
gauge group here again in agreement with \cite{KV:N=2}.

Determination of the $SU(2)\times SU(2)$ or $SU(3)$ points of enhanced
gauged symmetry is harder to analyze in this direct manner and we will
not pursue it here. It is clear that we should again reproduce the
results of \cite{KV:N=2} however.

As a second example we turn to one of the conjectured dual pairs of
\cite{AFIQ:chains} which has also been analyzed in
\cite{HM:alg}. Consider the \CY\ hypersurface of degree 84 in
$\P^4_{\{1,1,12,28,42\}}$. This is a K3 fibration where the generic
fibre is a hypersurface of degree 42 in $\P^3_{\{1,6,14,21\}}$. This K3
surface has Picard number 10 and the Picard lattice has intersection
form $E_8\oplus H$. This is most easily seen following the methods
explained in \cite{Mart:CCC}. Thus we have
\begin{equation}
  \Lambda_K \cong E_8\oplus H\oplus H.
\end{equation}

It is easy then to see that there will be dual heterotic strings with
an $E_8$ factor in the gauge group in agreement with
\cite{AFIQ:chains}. We can also say in more detail how to obtain this
gauge group. Let us label the simple roots of $E_8$ as follows:
\begin{equation}
\hbox{\unitlength=1mm\begin{picture}(42,14)(0,0)
  \multiput(0,3)(7,0){7}{\circle{1.8}}
  \multiput(0.9,3)(7,0){6}{\line(1,0){5.2}}
  \put(14,10){\circle{1.8}}
  \put(14,3.9){\line(0,1){5.2}}
  \put(-1.0,-1.0){\makebox(3.0,1.8)[b]{$e_2$}}
  \put(6.0,-1.0){\makebox(3.0,1.8)[b]{$e_3$}}
  \put(13.0,-1.0){\makebox(3.0,1.8)[b]{$e_4$}}
  \put(20.0,-1.0){\makebox(3.0,1.8)[b]{$e_5$}}
  \put(27.0,-1.0){\makebox(3.0,1.8)[b]{$e_6$}}
  \put(34.0,-1.0){\makebox(3.0,1.8)[b]{$e_7$}}
  \put(41.0,-1.0){\makebox(3.0,1.8)[b]{$e_8$}}
  \put(15.5,8.8){\makebox(3.0,1.8){$e_1$}}
\end{picture}}  \label{eq:E8}
\end{equation}
Each of these roots is associated with a rational curve in the K3
fibre. All the roots, with the exception of $e_4$, come from
blowing-up the quotient singularities of the ambient
$\P^4_{\{1,1,12,28,42\}}$. Thus we can blow these down to get a gauge
group $SU(2)\times SU(3)\times SU(5)$ while keeping the fibre at the
large radius limit. The curve corresponding to $e_4$ comes from the
hyperplane section from the ambient projective space however. To blow
this down we have to shrink down the whole K3 fibre to take us into
the realms of quantum geometry. Thus in order to obtain the full $E_8$
symmetry the fibre must be shrunk down to volume of order
$(\alpha^\prime)^2$.

Actually one can enhance the gauge group beyond
this using the $H$ factors in $\Lambda_K$. For example
$E_8\times SU(3)$ or $SU(10)$ should appear in the moduli space.

One can see that we can reproduce all of
the gauge groups for type IIA strings compactified on hypersurfaces
listed in \cite{AFIQ:chains}. Note also that any enhanced gauge groups
appearing in the dual pairs studied in \cite{FHSV:N=2,me:flower} can also
be recovered by the same methods.


\section{Comments}	\label{s:conc}

We have seen how enhanced gauge groups appearing on a type IIA string
compactified on a \CY\ manifold that can be seen perturbatively in the
dual heterotic string can be understood purely in terms of
string-string duality in six-dimensions. We used this fact to
reproduce all the currently known results about gauge groups from
conjectured dual pairs.

We should emphasize that we have not completed the problem of
understanding the appearance of nonabelian groups for type II strings
on \CY\ manifolds however. Firstly there can be parts of the gauge
group which cannot be understood perturbatively from either the type
II or the heterotic string point of view. Such groups have been
analyzed recently in six dimensions by using nonperturbative methods
for a heterotic string compactified on a K3 surface
\cite{W:small-i}. Similar effects must be expected for the case
considered in this letter.
It is tempting to conjecture how they will appear. We know from
\cite{AL:ubiq} that the vector multiplets that cannot be seen
perturbatively in the heterotic string arise from contributions to the
Picard group of $X$ from the degenerate fibres. We also have seen
above that curves of quotient singularities lead to nonabelian groups
in the type IIA string when this curve can be fibred over the base
$\P^1$. If we suppose that the appearance of nonabelian groups is a
result purely of singular curves and not whether they fibre properly
over the base $\P^1$ then we can consider the case that we pick up a
curve of singularities within the degenerate fibre. Thus, if we
have a type IIA string compactified on $X$ and $X$ is a K3 fibration
with curves of quotient singularities within the degenerate fibres,
then the dual heterotic string will have nonperturbatively enhanced
gauge groups as in \cite{W:small-i}. We should add however that one
may have to worry about what one means by the ``weak-coupling limit''
in which one actually sees these nonabelina gauge groups.
This should be investigated further.

Another aspect which we ignored above concerns hypermultiplets. We
have explored questions involving vector multiplets
only. Hypermultiplets can become massless when the gauge group gets
enhanced. This is essential for analysis of phase transitions as
discussed in \cite{GMS:con,FHSV:N=2,me:flower}. On a related point,
gauge groups can also become enhanced as we move about in the moduli
space of hypermultiplets.
The $D$-brane picture should be of help here. We may also need to
worry about discrete R-R degrees of freedom since precisely these
issues concerning hypermultiplets were sensitive to such effects in
the example of \cite{FHSV:N=2}.
Clearly we must address these problems before
the subject of enhanced gauge symmetries on \CY\ manifolds is
completely understood.


\section*{Acknowledgements}

It is a pleasure to thank D. Morrison
for useful conversations.
The work of the author is supported by a grant from the National
Science Foundation.


\end{document}